%% file: paper_v9.tex
\documentclass[manyauthors,nocleardouble,COMPASS]{cernphprep}

\usepackage{bm}
\usepackage{color}

\usepackage{amssymb}
\usepackage{amsmath}
\usepackage{amsbsy}
\usepackage{times}
\usepackage{epsfig} 
\usepackage{colordvi}
\usepackage{graphicx}
\usepackage{cite}
\usepackage{hyperref}
\usepackage{multicol}
\usepackage{booktabs}
\usepackage{lineno}
\usepackage{lscape}

\RequirePackage[T1]{fontenc}

\newcommand{\rk}{R_{\rm K}}

\begin{document}


\begin{titlepage}

\PHnumber{2018-012v2}
\PHdate{\today}  
\title{K$^{-}$ over K$^{+}$ multiplicity ratio for kaons produced in DIS \\ 
with a large fraction of the virtual-photon energy}

\date{}
\Collaboration{The COMPASS Collaboration}
\ShortAuthor{The COMPASS Collaboration}

\begin{abstract}
\label{abstract}

The K$^{-}$ over K$^{+}$ multiplicity ratio is measured in deep-inelastic scattering, 
for the first time for kaons carrying a large fraction $z$ of the virtual-photon energy.
The data were obtained by the COMPASS collaboration using a 160 GeV muon beam and an isoscalar 
$^6$LiD target.
The regime of deep-inelastic scattering is ensured by requiring $Q^2>1$ (GeV/$c)^2$ 
for the photon virtuality and $W>5$ GeV/$c^2$ for the invariant mass of the produced hadronic system.  
Kaons are identified in the momentum range from 12 GeV/$c$ to 40 GeV/$c$, 
thereby restricting the range in Bjorken-$x$ to $0.01<x<0.40$. 
The $z$-dependence of the multiplicity ratio is studied for $z>0.75$. 
For very large values of $z$, $i.e.$ $z>0.8$, 
we observe the kaon multiplicity ratio to fall below 
the lower limits expected from calculations based on 
leading and next-to-leading order perturbative quantum chromodynamics. 
Also, the kaon multiplicity ratio shows a strong dependence on the missing 
mass of the single-kaon production process. This suggests that 
within the perturbative quantum chromodynamics formalism an additional correction may be required, 
which takes into account the phase space available for hadronisation.  
\end{abstract}

\vfill
\Submitted{(to be submitted to Phys. Lett. B)}

\end{titlepage}

{\pagestyle{empty}  
\input{Authors2018.tex}
\clearpage
}

\maketitle

\section{Introduction} \label{sec:int}

Quark fragmentation into hadrons is a process of fundamental nature. 
In perturbative quantum chromodynamics (pQCD), 
this process is effectively described by non-perturbative
objects called fragmentation functions (FFs). 
While these functions presently cannot be predicted by theory, 
their scale evolution is described by the DGLAP equations \cite{dglap}. 
In leading order (LO) pQCD, the FF $D_{\rm q}^{\rm h}$ represents a probability density,
which describes the scaled momentum distribution of a
hadron type h that is produced in the fragmentation of a quark with flavour q.
 
The cleanest way to access FFs is to study hadron production in 
single-inclusive annihilation, $e^+ + e^- \to {\rm h}+X$, 
where the remaining final state $X$ is not analysed. 
These studies have two disadvantages: $i)$ 
that only information  
about $D_{\rm q}^{\rm h}+D_{\bar{\rm q}}^{\rm h}$
is accessible, and $ii)$ without invoking model-dependent algorithms for quark-flavour tagging
only limited flavour separation is possible.
In contrast, the analysis of semi-inclusive measurements of 
deep-inelastic lepton-nucleon scattering (SIDIS) is advantageous 
in that q and $\bar{\rm q}$ can be accessed separately 
and full flavour separation is possible in principle. 
Here, the disadvantage is that in the pQCD description of a 
SIDIS measurement FFs appear convoluted with 
parton distribution functions (PDFs).

Recently, COMPASS reported results on charged-hadron, 
pion and kaon multiplicities obtained over a wide kinematic 
range~\cite{comp_pi, comp_K}. 
These results provide important input for 
phenomenological analyses of FFs. 
The pion multiplicities were found to be well described both in 
leading-order (LO) and next-to-leading order (NLO) pQCD, 
while this was not the case for kaon multiplicities. The region 
of large $z$ appears to be particularly problematic for kaons, 
as it was also observed in subsequent analyses~\cite{lss_pc} 
of the COMPASS multiplicities. 
Here, $z$ denotes the fraction 
of the virtual-photon energy carried by the produced hadron in the
target rest frame.

In this Letter, we present results on the K$^-$ over K$^+$ 
multiplicity ratio in the large-$z$ region, $i.e.$ for $z > 0.75$. 
Instead of studying multiplicities for K$^-$ and K$^+$ separately, 
their ratio $\rk$ is analysed as in this case most experimental 
systematic effects cancel. Similarly, the impact of theoretical uncertainties,
$e.g.$ scale uncertainties, is largely reduced in the ratio.
Also, while pQCD cannot predict values of
multiplicities, limits for certain multiplicity ratios can be 
predicted. 
The Letter is organised as follows:
in Section \ref{sec:th} various predictions
for  $\rk$  are discussed.
The experimental set-up and the data selection
are described in Section \ref{sec:expdata}.
The analysis method is presented in Section \ref{sec:ana},
followed by the discussion of the systematic uncertainties in
Section \ref{sec:sys}.
The results are presented and discussed in Section \ref{sec:res}.

\section{{Theoretical framework} and model expectations} \label{sec:th}

Hadrons of type h produced in a SIDIS measurement 
are commonly characterised by their relative abundance. The hadron multiplicity 
$M^{\rm h}$ is defined as the ratio of the SIDIS cross section for  
hadron type h to the cross section for an inclusive measurement of the deep-inelastic scattering process: 
\begin{equation}
\label{mult_def}
     \frac{{\rm d} M^{\rm h}(x,Q^2,z)}{{\rm d}z} =\frac{{\rm d}^3\sigma^{\rm h}(x,Q^2,z)/{\rm d}x {\rm d}
Q^2 {\rm d} z}{{\rm d}^2\sigma^{\rm DIS}(x,Q^2)/{\rm d}x {\rm d} Q^2}.
\end{equation}
Here, $Q^2$ is the virtuality of the photon mediating the lepton-nucleon scattering process and 
$x$ denotes the Bjorken scaling variable. 
Within the standard factorisation approach of pQCD \cite{nlo_form,dsv}, 
$\sigma^{\rm DIS}$ can be written as a sum over parton types, 
in which for a given parton type $a$ the
respective PDF is convoluted with the lepton-parton hard-scattering cross section.
For $\sigma^{\rm h}$ in the current fragmentation region,  the sum 
contains an additional convolution with the fragmentation function of the produced parton.
The rather complicated NLO expressions for these cross sections can be found $e.g.$ in Ref.~\cite{dsv}.
Below, we will use only pQCD LO expressions for the cross section, 
while later for the presentation of results also multiplicity calculations 
obtained using NLO expressions will be shown.
It is important to note that in the SIDIS factorisation approach 
the only ingredients that depend on the nucleon type are the nucleon PDFs, 
while the fragmentation functions depend neither 
on the nucleon type nor on $x$. 
In the LO approximation for the multiplicity, 
the sum over parton species $a= {\rm q},  \bar{\rm q}$ does not contain convolutions 
but only simple products of PDFs $f_a(x,Q^2)$, 
weighted by the square of the electric charge $e_a$ of the quark
expressed in units of elementary charge, and FFs $D_a^{\rm h}(z,Q^2)$:

\begin{equation} 
\label{mult_LO}
    \frac{{\rm d} M^{\rm h}(x,Q^2,z)}{{\rm d}z}=\frac{\sum_a{e_a^2f_a(x,Q^2)D_a^{\rm h}(z,Q^2)}}{\sum_a{e_a^2f_a(x,Q^2)}}.
\end{equation}
For a deuteron target, the charged-kaon multiplicity ratio in LO pQCD reads as follows:

\begin{equation} \label{eq:main0}
\rk (x,Q^2,z)= \frac{{\rm d}M^{{\rm K}^{-}}\!(x,Q^2,z)/{\rm d}z}{{\rm d} M^{{\rm K}^{+}}\!(x,Q^2,z)/{\rm d}z}=
\frac{4(\bar{\rm u}+ \bar{\rm d}) D_{\rm fav} +({\rm 5u + 5d+ \bar{u}+ \bar{d} + 2\bar{s}}) D_{\rm unf} +  2{\rm s}D_{\rm str} }
     {4( {\rm u + d}) D_{\rm fav} +  (5 \bar{\rm u} + 5 \bar{\rm d} + {\rm u} + {\rm d}+ 2 {\rm s}) D_{\rm unf} + 2 \bar{\rm s}D_{\rm str} }.
\end{equation}

Here, u, $\bar{\rm u}$, d, $\bar{\rm d}$, s, $\bar{\rm s}$ 
denote the PDFs in the proton for different quark flavours. 
Their dependences on $x$ and $Q^2$ are omitted for brevity. 
The symbols $D_{\rm fav}$, $D_{\rm unf}$ and $D_{\rm str}$ denote favoured, 
unfavoured, and strange-quark fragmentation functions respectively, 
which are given by $D_{\rm fav}= D_{\rm u}^{\rm K^{+}} = D_{\bar{\rm u}}^{\rm K^{-}}$, 
$D_{\rm unf}= D_{\bar{\rm u}}^{\rm K^{+}} = D_{\rm d}^{\rm K^{+}}= D_{\bar{\rm d}}^{\rm K^{+}} = 
D_{\rm s}^{\rm K^{+}}$  and their charge conjugate,
and $D_{\rm str}= D_{\bar{\rm s}}^{\rm K^{+}} = D_{\rm s}^{\rm K^{-}}$. 
Their dependences on $z$ and $Q^2$ are omitted. 
Accordingly, also the dependence of $\rk$
on $x$, $Q^2$ and $z$ are omitted.
Presently, existing data do not allow one to distinguish
between different functions $D_{\rm unf}$ 
for different quark flavours. 
However, it is expected that $D_{\rm unf}$ is small in the large-$z$ region, 
and this expectation is indeed confirmed in pQCD fits already
at moderate values of $z$, $i.e.$ $z \approx 0.5$, see e.g. Refs. \cite{dss01, dss02}.
When neglecting $D_{\rm unf}$, Eq.~(\ref{eq:main0}) simplifies to
\begin{equation}
\rk = 
 \frac{4(\bar{\rm u}+ \bar{\rm d} )D_{\rm fav} +  2{\rm s} D_{\rm str}}
      { 4({\rm u+ d}) D_{\rm fav} + 2 \bar{\rm s} D_{\rm str}}.
\end{equation}
It is expected that $D_{\rm str}>D_{\rm fav}>0$, and 
therefore the positive terms s$D_{\rm str}$ and $\bar{\rm s}D_{\rm str}$ may be of some importance. 
Still, in order to calculate a lower limit for $\rk$, 
these terms can be neglected under the assumption that ${\rm s}=\bar{\rm s}$,
which leads to
\begin{equation} \label{eq:final}
\rk > \frac{\bar{\rm u} + \bar{\rm d}}{\rm u+d}.
\end{equation} 

The analysis described below is performed using two bins in $x$,
$i.e.$ $x<0.05$ with $\langle x \rangle = 0.03$, $\langle Q^2 \rangle = 1.6$
(GeV/$c)^2$ and $x>0.05$ with $\langle x \rangle = 0.094$, $\langle Q^2 \rangle = 4.8$
(GeV/$c)^2$. Whenever sufficient, only the first $x$-bin is used in the
discussion.

The evaluation of Eq.~(\ref{eq:final})
for $x=0.03$ and $Q^2=1.6$ (GeV/$c)^2$ yields a lower
limit of $0.469 \pm 0.015$ when using the MSTW08 LO PDFs \cite{pdf_mstw08}.
In NLO the limit given by Eq.~(\ref{eq:final}) receives corrections on the level of
$\sim \alpha_{S}/2\pi$. Using the  MMHT14 NLO PDF set \cite{pdf_mmht14},
the ratio $(\bar{\rm u} + \bar{\rm d})/({\rm u+d})$ is $0.440\pm  0.023$,
but according to our calculation the lower limit is about 15\% lower than this limit
\footnote{From the formalism given in \cite{nlo_form},
it follows that in the NLO cross-section
formula for hadron production, for each quark flavour there are six additional terms
besides the q$D^{\rm h}_{\rm q}$ term. These terms include convolution integrals of
of PDF, FFs and the so-called coefficient functions.
We found that four convolution integrals can effectively be neglected at high $z$, 
and only two that are related 
to convolutions of $C^1_{\rm qq}$ and $C^1_{\rm qg}$ have an important impact on the final results. 
The term related to $C^1_{\rm qq}$ alone would
lead to an increase of $\rk$ above the limit given by Eq.~(\ref{eq:final}).
In contrast, the term related to $C^1_{\rm qg}$, although appearing 
in a symmetric form in numerator and denominator, 
is negative, so that the lower limit of $\rk$ falls below that given by Eq.~(\ref{eq:final}).
We note that $D_{\rm fav}$ or its convolution appears always in all relevant terms.
Its choice hence appears to be rather irrelevant for the final result, as it largely cancels in
the predicted lower limit for $\rk$ at NLO.}.

We note that because of the large uncertainties of ${\rm s}, \bar{\rm s}$ and $D_{\rm str}$,
reasonable uncertainties are presently calculable only for the lower limits of $\rk$, 
and not for $\rk$ itself. These uncertainties amount to about 3\% for LO and about 
6\% for NLO predictions. 
In both cases the uncertainty of the $(\bar{\rm u}+ \bar{\rm d})/({\rm u+d})$ ratio dominates,
while in NLO also uncertainties of the gluon PDF play some role. The choice of
FFs has negligible impact on LO or NLO calculations of the lower $\rk$ limit. 
The actual predictions for $\rk$ based on
DSS~\cite{dss01} at LO accuracy and DEHSS17~\cite{dss02}
at NLO accuracy are larger than the lower limits for $\rk$,
which is expected as in the above calculation of lower limits
the strange-quark contribution to kaon fragmentation was neglected.
It was verified that when using more recent PDF sets
($e.g.$ NNPDF30 at LO and NLO accuracy~\cite{pdf_nnpdf}),
the $\rk$ values increase by about 10\% for all cases that were
discussed above. 
Hence our choice of the MSTW08 LO and MMHT14 NLO PDFs sets 
leads to a rather conservative estimation of the lower limit on $\rk$.

In the LEPTO event generator
\footnote{
LEPTO 6.5, with JETSET 7.4 and fragmentation tuning from Ref. \cite{comp-allpt}.} \cite{lepto}.
another factorisation ansatz is used 
\begin{equation}
\label{mult_LEPTO}
    \frac{{\rm d} M^{\rm h}(x,Q^2,z)}{{\rm d}z}=\frac{ \sum_a{e_a^2f_a(x,Q^2)H_{a/N}^{\rm h}(x,z,Q^2)}}{\sum_a{e_a^2f_a(x,Q^2)}}.
\end{equation}
Here, $H^{\rm h}_{a/N}(x,z,Q^2)$ describes the production of a hadron h in the hadronisation of a string 
that is formed by the struck quark and the target remnant. 
In contrast to the pQCD approach, this hadronisation function depends not only on quark and 
hadron types and on $z$ but also on the type of the target nucleon 
and on $x$, see Ref.~\cite{Aram} for more details. 
We note that in this approach also the conservation of the overall quantum numbers as well as
momentum conservation are taken into account, which is not the case for the pQCD approach. 
The LEPTO prediction for $\rk$, about 0.52, lies above the LO limit given by Eq.~(\ref{eq:final}). 
However, for $z>0.97$ it undershoots this limit. 
This appears plausible as for $z$ approaching unity K$^+$ can be produced in the process 
$\mu {\rm p} \rightarrow \mu {\rm K}^+ \Lambda^{0}$, while a
similar process to produce K$^-$ is forbidden because of baryon number
conservation.

In recent years, several theory developments were performed that 
can potentially impact the theory predictions for the high-$z$ region. 
In Ref. \cite{WVog_res} for example, the authors studied the impact of 
threshold-logarithm resummations in the high-$z$ region and found a large impact.
In the case of $\pi^{-}$ production, the predicted cross section can be larger
by a factor of two.
When considering the lower limit for $\rk$,
the resummation corrections for K$^-$ and K$^+$ are largely 
proportional to the PDF densities $\bar{\rm u}+\bar{\rm d}$ and u+d, respectively.
Therefore, the $\rk$ predictions 
including these resummation corrections would be even closer to the expectations 
given by Eq.~(\ref{eq:final}) than the NLO predictions shown below without including these 
corrections. An interesting work related to hadron-mass corrections \cite{TMC_old}
was originally criticised in Ref. \cite{TMC_ChL}, but the discussion is ongoing \cite{TMC_new}. 
The approach discussed in this work allows one to obtain a value of $\rk$ below the limits discussed above. 
However, this approach seems to go beyond the standard factorisation theorem 
and corrections to $D_{\rm q}^{\rm h}$ are needed, which depend on the type of target nucleon 
and produced hadron h. 
There were also other developments, $e.g.$ Refs. \cite{Ref2_1, Ref2_2, Ref2_3},
which are very important for a better understanding of the hadronisation process. Still,
they appear to not effectively impact the predictions 
for $\rk$ in the high-$z$ region at COMPASS kinematics.

\section{Experimental set-up and data selection} \label{sec:expdata}

The data were taken in 2006 using a $\mu^+$ beam  delivered by the M2 beam line of the CERN SPS.
The beam momentum was $160$ GeV/$c$ with a spread of $\pm$ 5\%.
The solid-state $^6$LiD target is considered to be purely isoscalar, neglecting 
the 0.2\% excess of neutrons over protons due to
the presence of additional material in the target ($^3$He and $^7$Li).
The target was longitudinally polarised but in the present analysis
the data are averaged over the target polarisation, 
which leads to an effectively vanishing target polarisation on a level of better than 1\%. 
The COMPASS two-stage spectrometer has a polar angle acceptance of $\pm$180 mrad,
and it is capable of detecting charged particles with momenta above 0.5 GeV/$c$.
The ring-imaging Cherenkov detector (RICH) was used to identify pions, kaons and protons. 
Its radiator volume was filled with C$_4$F$_{10}$ 
leading to a threshold for pion, kaon and proton identification of about 3 GeV/$c$, 9 GeV/$c$ and 
18 GeV/$c$ respectively.
Efficient pion and kaon separation is possible with high purity for
momenta between 12 GeV/$c$ and 40 GeV/$c$.
Two trigger types were used in the analysis.
The ``inclusive'' trigger was based on a signal from a combination of hodoscope signals 
from the scattered muon. The ``semi-inclusive'' trigger required an energy deposition in one of
the hadron calorimeters.
The experimental set-up is described in more detail in Ref. \cite{nim}.

The data selection criteria are kept similar
to those used in the recently published analysis \cite{comp_K}, whenever possible.
The kinematic domain $Q^2 > 1$ (GeV/$c)^2$ and $W > 5$ GeV/$c^2$  is selected, 
thereby restricting the analysis to the region 
of deep inelastic scattering where pQCD can be applied.
For small values of $y$, $i.e.$ the fraction of the incoming muon energy carried by the virtual photon, 
the momentum resolution is degraded. 
In order to exclude this region, $y$ is required to have a minimum value of 0.1.
The aim of this analysis is to study kaon production in SIDIS
for kaons carrying a large fraction $z$ of the virtual-photon energy, hence it 
is restricted to $z>0.75$.
Using the above given momentum range for efficient kaon identification 
together with the large-$z$ requirement in this analysis leads to an effective upper limit for $y$ of 0.35.

The kaon multiplicities \mbox{$M^{\rm K}$($x$, $Q^2$, $z$)} 
are determined from the
kaon yields $N^{\rm K}$ normalised by the number of DIS events, 
$N^{\rm DIS}$, and divided by the acceptance correction $A^{\rm K}(x,Q^2,z)$:
\begin{equation} \label{eq:ExpMul}
\frac{\text{d}M^{\rm K}(x,Q^2,z)}{\text{d}z} =
\frac{1}{N^{\rm DIS}(x,Q^2)}\frac{\text{d}N^{\rm K}(x,Q^2,z)}{\text{d}z} \frac{1}{A^{\rm K}(x,Q^2,z)}\,.
\end{equation}
Note that in this work ``semi-inclusive'' triggers can be used
because a bias free determination of $N^{\rm DIS}$ is not needed,
as the latter cancels in $\rk$.

\begin{figure}
\centerline{\includegraphics[clip,width=1.0\textwidth]{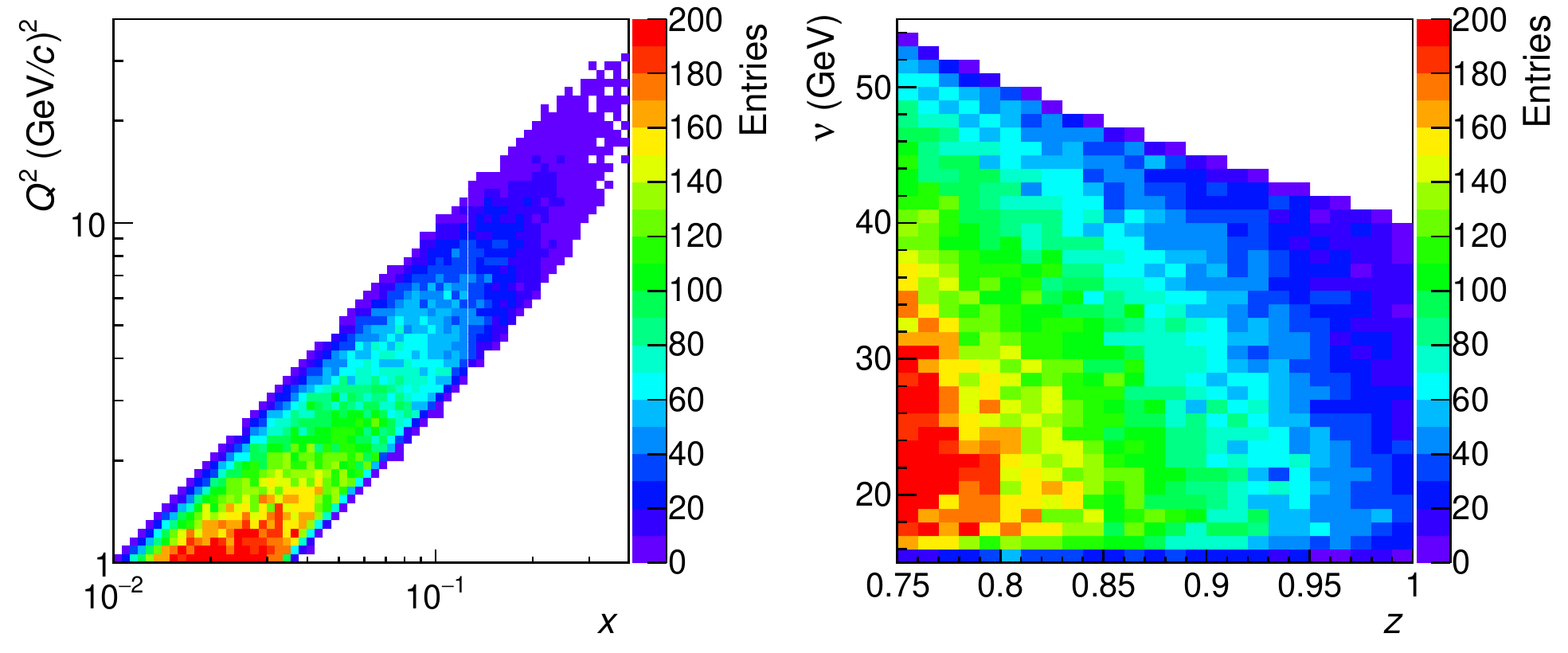}}
\caption{Acceptance-uncorrected distributions of selected events in the ($Q^2$, $x$) plane and in the ($\nu$, $z$) plane.}
\label{fig:kine}
\end{figure}

All data taken in 2006 are used in the analysis;
altogether about 64000 charged kaons are available in the region $z > 0.75$.
Examples of acceptance-uncorrected distributions of selected events are presented in Fig. \ref{fig:kine} 
in the ($x$, $Q^2$) 
and ($\nu$, $z$) planes.
Here, $\nu$ is the energy of the virtual photon in the laboratory frame.

\section{Analysis method} \label{sec:ana}

The analysis is performed in two $x$-bins, below and above $x=0.05$, 
as already mentioned in Section \ref{sec:th}.
In each $x$-bin, five bins
are used in the reconstructed $z$ variable ($z_{\rm rec}$) 
with the bin limits
0.75, 0.80, 0.85, 0.90, 0.95, 1.05.
Since the RICH performance depend upon the 
momentum of the identified kaon, we 
also study $\rk$ in bins of this variable 
using the bin limits  12 GeV/$c$, 16 GeV/$c$, 20 GeV/$c$, 25 GeV/$c$, 30 GeV/$c$, 35 GeV/$c$, 40 GeV/$c$.
Note that in this way 
the $\nu$ dependence of $\rk$ is studied implicitly 
and that the results are also given as a function of $\nu$ 
in these kaon-momentum bins.

In order to determine the multiplicity ratio $\rk$ from the raw yield of
K$^-$ and K$^+$ mesons, several correction factors have to be taken into account.
First, the number of identified kaons is corrected for the RICH efficiencies.
Based on studies of $\phi \rightarrow {\rm K}^+ {\rm K}^-$ decays,
where the $\phi$  meson was produced
in a DIS process, the efficiency ratio for the two charges is found to be $1.002 \pm 0.012$.
Such a simple ``unfolding'' procedure can be followed
because a strict selection of kaons is made,
so that the probabilities of misidentification of pion and proton as kaon 
can be assumed to be zero (possible remaining misidentification
probabilities are discussed in Section~\ref{sec:sys}).

\begin{figure}
\centerline{\includegraphics[clip,width=0.49\textwidth]{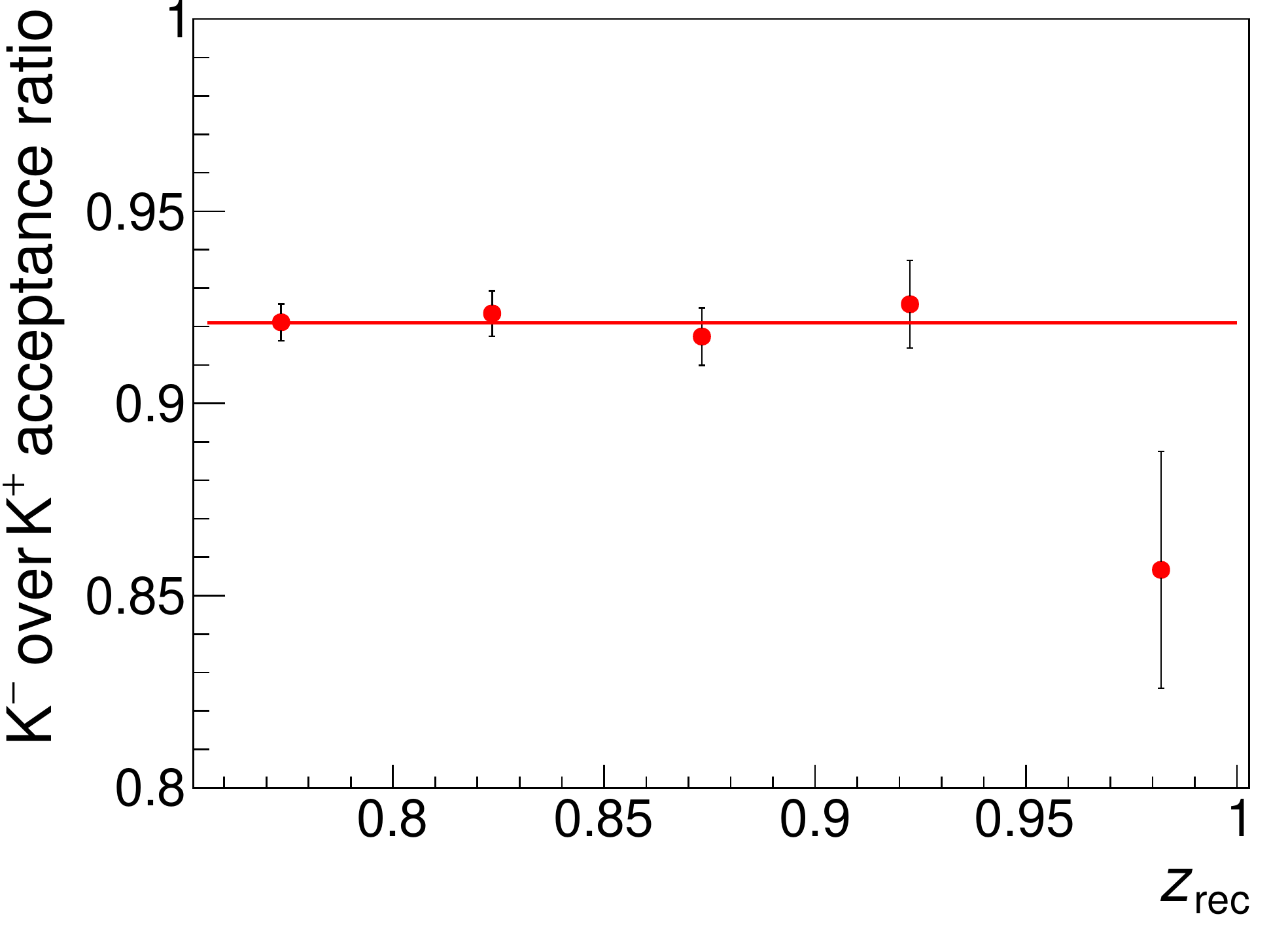}
            \includegraphics[clip,width=0.49\textwidth]{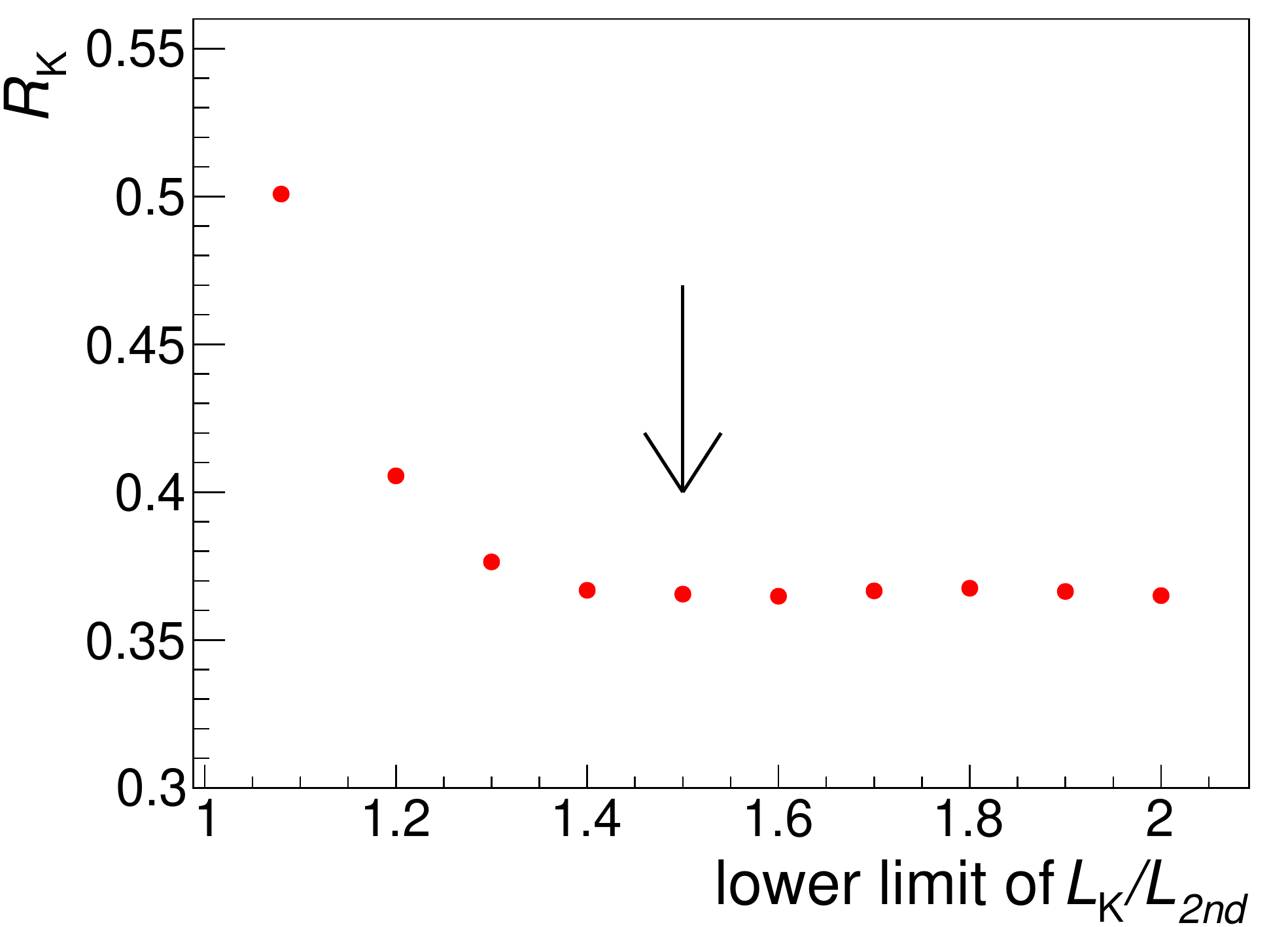}}
\caption{Left: The K$^-$ over K$^+$ acceptance ratio in the first $x$-bin, $i.e.$ $x<0.05$, as a
function of the reconstructed $z$ variable, as obtained from a Monte Carlo simulation. 
Right: The charged-kaon multiplicity ratio in the first $x$-bin, 
as a function of the lower limit of the RICH likelihood ratio 
for kaons with momenta between 35 GeV/$c$ and 40 GeV/$c$.
The arrow marks the value used in the analysis (see text for more details).} 
\label{fig:accK}
\end{figure}

The acceptance correction factors $A^{\rm K}$ for the two kaon charges 
are determined using Monte Carlo simulations. 
In the previous COMPASS analysis \cite{comp_K}, a simple unfolding 
method was used to determine these factors. 
For a given kinematic bin in $(x,y,z)$,
the acceptance was calculated 
as the ratio of the number of reconstructed events to that of generated ones. 
For a given event, reconstructed variables were used to count reconstructed events 
and generated variables to count generated events. 
In order to account for the strong $z$-dependence of the multiplicity in the large-$z$ region, 
in this analysis the acceptance is unfolded as in Ref. \cite{comp_K}
for $x$ and $Q^2$ but not for $z$. 
Various methods for $z$ unfolding were investigated in detailed studies,
see appendix A for an example.
The results presented in this Letter are obtained
using the simplest version of $z$ unfolding,
$i.e.$ unfolding only the dependence of $\rk$
on $z_{\rm corr}$.
Here, $z_{\rm corr}$ denotes the reconstructed value of $z$ in the experiment,
corrected by the average difference between the generated and reconstructed values of $z$,
where the latter are determined by Monte Carlo simulations.
In the left panel of Fig.~\ref{fig:accK}, the K$^-$ over K$^+$ acceptance ratio 
obtained from $x$ and $Q^2$ unfolding is shown as a function 
of the reconstructed $z$-variable in the first $x$-bin. 
It appears to be independent of $z$ within statistical uncertainties 
and has a value of $0.921 \pm 0.004$ in the first $x$-bin and $0.969\pm 0.010$ in the second $x$-bin.

The contamination by decay products of diffractively
produced vector mesons is estimated using HEPGEN \cite{hepgen} and found 
to be negligible, see Fig.~2 in \cite{comp_K}. 
Only $\phi$ decays are simulated there since heavier
vector mesons have cross sections smaller by a factor of about 10  
and decay mostly in multi-body channels, which  
results in even smaller probabilities to produce kaons at large $z$.

The measured cross sections have to be corrected for radiative effects in order to obtain $\sigma^{\rm DIS}$ and $\sigma^{\rm h}$. 
Since, $y< 0.35$ holds as explained above, the size of radiative corrections is expected to be small.
In any case, $\sigma^{\rm DIS}$ cancels in $\rk$ 
and in the TERAD code \cite{terad} used in COMPASS analyses 
the relative radiative correction is the same for
K$^+$ and K$^-$, so that it also cancels in the ratio.

\section{Systematic studies} \label{sec:sys}

The charged-kaon multiplicity ratios measured in this analysis 
are found to agree with the results of the previous analysis
\cite{comp_K} in the overlap region of the $z$-ranges used in these
two analyses $(0.75<z<0.85)$.
Results derived from data that were obtained using different 
triggers are found to agree with one another within 2\%.

The most important correction factor is the K$^-$ over K$^+$ acceptance ratio,
which for the first $x$-bin is $0.921\pm 0.004$, 
as obtained using Monte Carlo data.
The COMPASS spectrometer is designed to be almost charge symmetric. 
In the case of pions, 
the acceptance ratio obtained from Monte Carlo simulations is $0.991\pm 0.003$, $i.e.$ very close to unity. 
In contrast, 
the acceptance ratio of kaons obtained from Monte Carlo is found to be significantly below unity. 
This difference between K$^-$ and K$^+$ yields
is caused by the non-negligible thickness
of the COMPASS target,
which amounts to about 50\% of 
a hadron interaction length, combined with a considerably larger absorption cross
section for interactions of negative kaons compared to positive ones, 
see $e.g.$ the results on the K$^\pm$-deuteron cross section in Ref. \cite{pdg}. 
Depending 
on the longitudinal position of the primary interaction point 
$Z_{\rm vtx}$, the produced kaons traverse a varying thickness of the 
material contained in the
120 cm long target.  
As a result, more negative than positive kaons are absorbed 
when the interaction took place at the beginning of the target as compared to
an interaction at the end of the target. It is verified that 
once the acceptance correction was applied, the obtained $\rk$ ratio is flat as a function of $Z_{\rm vtx}$.
For the K$^-$ over K$^+$ acceptance ratio a 2\% systematic uncertainty is used; this value is dominated 
by possible trigger-dependent variations of the multiplicities mentioned in the previous paragraph.

The stability of
$\rk$ is tested on data using several variables
that are defined in the spectrometer coordinate system.
The most sensitive one is the azimuthal angle $\phi$ 
of the produced kaon.
The direction $\phi=0$ lies in the bending plane 
of the dipole magnets and points towards the side, to which positive 
particles are bent. Correspondingly, the direction
$\phi=\pi/2$ points towards the top of the spectrometer. 
In certain cases the charged-kaon multiplicity ratio is found to vary 
by up to 25\%, with particularly small values close to a peak at $\phi=0$.
This observation is accounted for by a systematic uncertainty that is taken 
as the difference between the multiplicity ratio measured over the full $\phi$-range 
and the one measured for $|\phi| > 0.5$.
Typically, the relative uncertainty related to this $\phi$-dependence 
ranges between 3\% and 11\%, which makes it the dominant
systematic uncertainty. Note that the values of this systematic 
uncertainty for different bins in $z$ are strongly correlated, with
a correlation coefficient of about 0.8.

Further systematic uncertainties may arise from the RICH identification procedure. 
The K$^-$ over K$^+$ efficiency ratio 
is expected to be close to
unity since the RICH detector is situated behind a dipole magnet of relatively weak bending power.
Additional studies were performed on data concerning misidentification probabilities
of pions and protons being identified as kaons by varying the ratio of the kaon likelihood, 
which is the largest of all likelihoods in the selected sample, to
the next-to-largest likelihood hypothesis, $L_{\rm K}/L_{2nd}$.
The behaviour of $\rk$ as a function of the lower limit for $L_{\rm K}/L_{2nd}$ is 
shown in the right panel of Fig.~\ref{fig:accK}
for kaon candidates with momenta between 35 GeV/$c$ and 40 GeV/$c$.
The constraint $L_{\rm K}/L_{2nd} > 1.5$ is used in the present analysis.
From these studies,
the systematic uncertainty of the RICH unfolding procedure of about 3\%. 
It corresponds to
the difference in $\rk$ calculated from the final sample and the one, 
in which a non-zero $\pi$ contamination is detected.

As the COMPASS muon beam is (naturally) polarised with  an average
polarisation of $-0.80 \pm 0.04$, a spin-dependent contribution to 
the 
total lepton-nucleon cross section
cannot be neglected a priori. This contribution 
is proportional to $\sin{\phi_{\rm h}}$ and expected to be smaller
than the spin-independent one, which is proportional to 
$\cos{\phi_{\rm h}}$ and $\cos{2 \phi_{\rm h}}$ \cite{comp_azi}.
Here, $\phi_{\rm h}$ denotes the azimuthal angle between the lepton-scattering plane 
and the hadron-production plane in the 
centre-of-mass frame of virtual photon and nucleon.
Studies performed for previous COMPASS measurements~\cite{comp_pi, comp_K} 
show that these effects can be neglected 
when using $\phi_{\rm h}$-integrated multiplicities, as it is done in this analysis. 

Altogether, the total relative systematic uncertainty on $\rk$ is found to
range between 5\% and 12\% depending upon the $z$-bin. 
The systematic uncertainties in different
$z$-bins are highly correlated, $i.e.$ the correlation coefficient is estimated to
vary between 0.7 and 0.8.

\section{Results and discussion} \label{sec:res}

\begin{figure}
\centerline{\includegraphics[clip,width=0.9\textwidth]{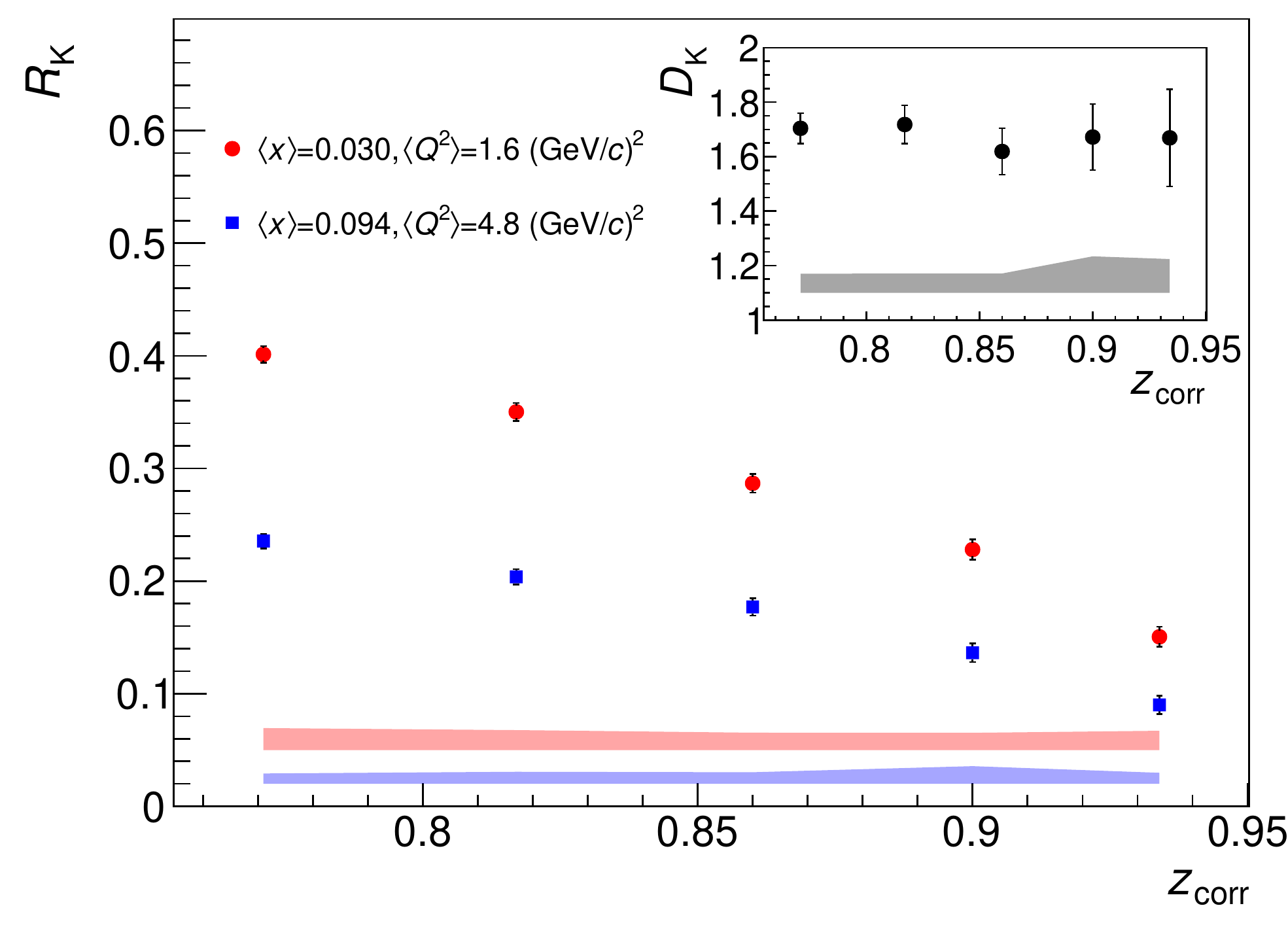}}
\caption{Results on $\rk$ as a function
of $z_{\rm corr}$ for the two $x$-bins. The insert shows the double ratio $D_{\rm K}$ 
that is the ratio of $\rk$ in the first $x$-bin over $\rk$ in the second $x$-bin. 
Statistical uncertainties are shown by error bars, systematic uncertainties by the shaded bands at the bottom.}
\label{fig:res1}
\end{figure}

In Table \ref{tab:res0}, the results on the charged-kaon multiplicity ratio 
$\rk$ are presented in bins of the reconstructed $z$ variable for the two $x$-bins. 
The measured $z$-dependence of $\rk$ can be fitted 
in both $x$-bins by simple functional forms, $e.g.$ $ 
\propto (1-z)^{\beta}$, $\beta=0.71\pm 0.03$. 
Dividing in every $z$-bin the value of the ratio measured in the first $x$-bin 
by the one measured in the second $x$-bin, a ``double ratio'' $D_{\rm K}=R_{\rm K}(x<0.05)/R_{\rm K}(x>0.05)$
is formed that appears to be 
constant over all the measured $z$-range with a value $D_{\rm K}=1.68\pm 0.04_{\rm stat.}\pm0.06_{\rm syst.}$. 
It is interesting to note that the measured value agrees within uncertainties 
with $D_{\rm K}$ calculated using the LO MSTW08L PDF set, $i.e.$ $1.56 \pm 0.07$.
In Fig.~\ref{fig:res1}, $\rk$ is shown 
as a function of $z_{\rm corr}$
for the two $x$-bins, as well as $D_{\rm K}$ in the inset of the figure. 
As both data and LO pQCD calculation exhibit the same $z$-dependence  
when comparing the charged-kaon multiplicity ratios in the two $x$-bins, 
in what follows we concentrate only on the first $x$-bin, $i.e.$ $x<0.05$. 
Still, the conclusions presented in the remaining part of the Letter are valid for both $x$-bins.

\begin{figure}
\centerline{\includegraphics[clip,width=0.9\textwidth]{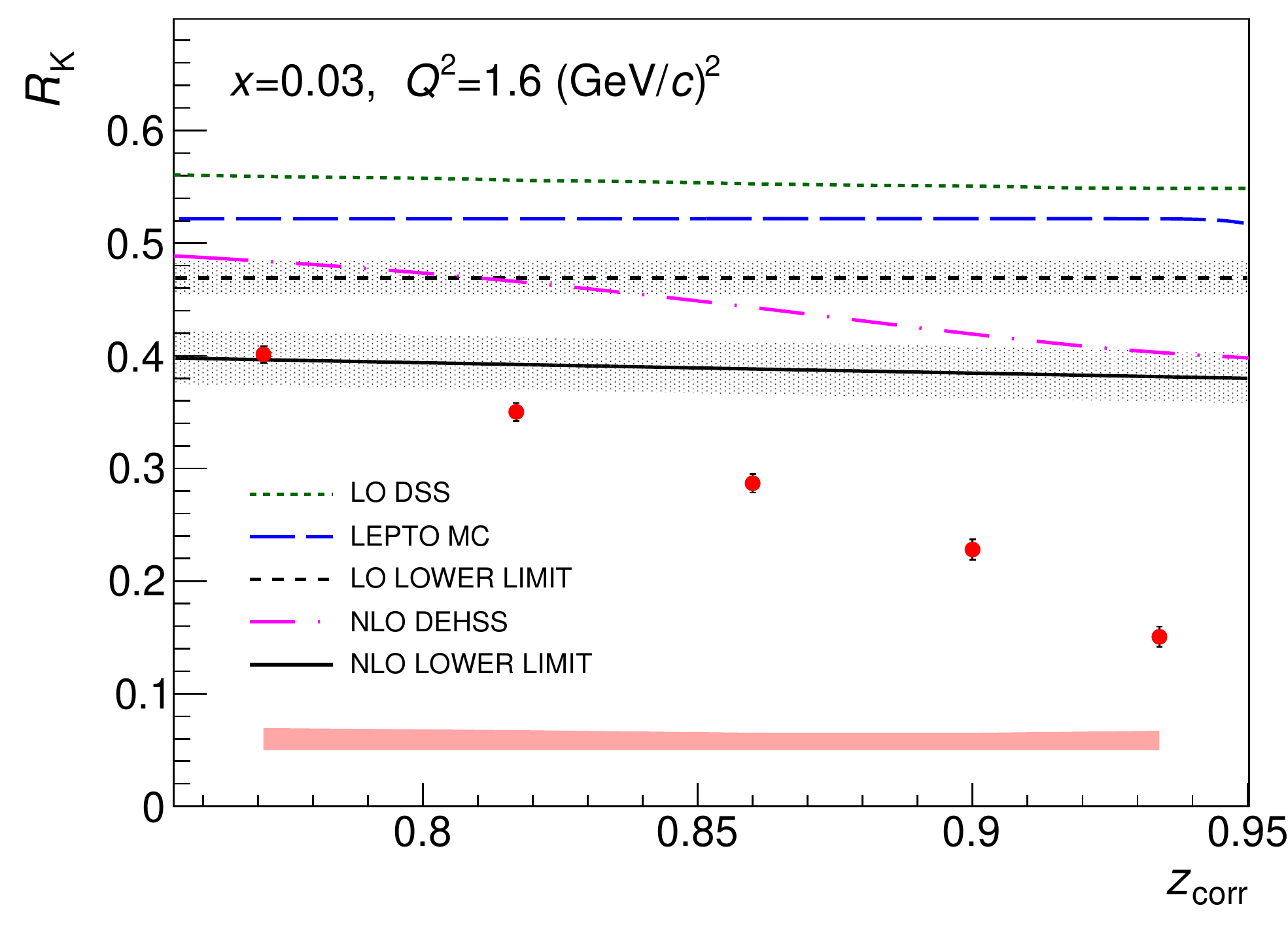}}
\caption{Comparison of $\rk$ in the first $x$-bin with predictions discussed 
in Section \ref{sec:th}. 
The systematic uncertainties of the data points are
indicated by the shaded band at the bottom of the figure.
The shaded bands around the (N)LO lower limits 
indicate their uncertainties. } 
\label{fig:res2}
\end{figure}

In Fig.~\ref{fig:res2}, the present results on $\rk$ 
in the first $x$-bin are compared with the expectations from LO and NLO pQCD calculations 
and with the predictions obtained using the LEPTO event generator,
which were all discussed in Section \ref{sec:th}. 
For completeness, we note that in the second $x$-bin
the typical $\rk$  predictions are about 1.5--1.6 times smaller than in the first $x$-bin.
It is observed that with increasing $z$
the values of $\rk$ are increasingly undershooting the expectations from LO and NLO calculations. 
The discrepancy between the COMPASS results and the NLO predictions reaches a factor 
of about 2.5 at the largest value of $z$. 
As the difference between the lower limit in LO and the NLO DEHSS 
prediction obtained under the assumption $D_{\rm str}=0$ is never larger than ~20\%, 
it is very unlikely that any prediction obtained at NNLO would be able to 
account for such a large discrepancy.

As already mentioned in Sect.~\ref{sec:th}, the presented pQCD calculations rely 
on the factorisation ansatz 
${\rm d}^3\sigma^{\rm h}(x,Q^2,z)/{\rm d}x {\rm d} Q^2 {\rm d}z \propto \sum_a{e_a^2f_a(x,Q^2)D_a^{\rm h}(z,Q^2)}$.
If this ansatz would not be applicable at COMPASS energies for large values of $z$, 
it may be incapable to describe the behaviour of kaon multiplicities in this kinematic region. 
This pQCD ansatz does not include higher-twist terms, which are  
proportional to powers $1/Q^2$,  
so that the respective correction should be smaller by a factor of about three 
in the second $x$-bin compared to the first $x$-bin. 
However, the discrepancy between COMPASS results and both LO and NLO predictions 
is observed to be the same in the two $x$-bins within experimental uncertainties. 
The observed discrepancy cannot be explained by 
the threshold resummations from Ref. \cite{WVog_res}, as discussed in Section \ref{sec:th}. 
The usage of DSS fragmentation functions \cite{dss01} in the LO ansatz 
presented in Ref. \cite{TMC_old} leads to a decrease of the $\rk$ prediction 
by about 25\% in the last $z$ bin.
It is thus not enough to account for the observed discrepancy. 
However, larger changes could be obtained if FFs decrease to zero faster than
expected in the DSS parametrisation. 
It is worth noting that in the LEPTO event generator
a different factorisation approach is used, which is based on string hadronisation.
However, it does not describe the data at high $z$,
in spite of its considerably higher flexibility
in comparison to the pQCD approach.
Perhaps a special tuning of certain string fragmentation parameters,
for example those governing low-mass string hadronisation,
would lead to a better description of the data.

In the analysis we assume that there is no contamination by
decay products of vector mesons or by
pions that were misidentified as kaons. 
Note that if these assumptions should not hold, the corrected
$\rk$ values would be further decreased
with respect to the results presented in this Letter,
$i.e.$ the disagreement with pQCD expectations would be
even stronger.

\begin{table}[h!]
\begin{center}
\caption{Extracted values of $\rk$,
bin limits of $z$ $(z_{\rm min}, z_{\rm max})$, and the
averages values of $x$, $Q^2$, $z_{\rm rec}$ and $z_{\rm corr}$ in first (upper part) and second
(lower part) $x$-bin.}
\begin{tabular}{ c|c|c|c|c|c|c|c }
bin &  $x $ & $ Q^2 $ (GeV/$c)^2$   & $z_{\rm min}$  & $z_{\rm max}$ & $ z_{\rm rec} $ &  $ z_{\rm corr} $   & $\rk \pm \delta R_{\rm K, \, stat.} \pm \delta R_{\rm K, \, syst.}$ \\ \hline
1 & 0.030& 1.7&  0.75 &0.80 & 0.774 & 0.771    & $0.401 \pm 0.007 \pm 0.019$ \\
2 & 0.030& 1.6&  0.80 &0.85 & 0.824 & 0.817    & $0.350 \pm 0.008 \pm 0.018$ \\
3 & 0.031& 1.6&  0.85 &0.90 & 0.873 & 0.860    & $0.287 \pm 0.008 \pm 0.015$ \\
4 & 0.031& 1.6&  0.90 &0.95 & 0.923 & 0.900    & $0.228 \pm 0.009 \pm 0.015$ \\
5 & 0.032& 1.5&  0.95 &1.05 & 0.982 & 0.934    & $0.150 \pm 0.009 \pm 0.017$ \\ \hline
$1^{'}$& 0.094& 5.1&  0.75 &0.80 & 0.774 & 0.771    & $0.235 \pm 0.007 \pm 0.009$\\
$2^{'}$& 0.094& 4.8&  0.80 &0.85 & 0.824 & 0.817    & $0.204 \pm 0.007 \pm 0.011$\\
$3^{'}$& 0.093& 4.6&  0.85 &0.90 & 0.873 & 0.860    & $0.177 \pm 0.008 \pm 0.010$\\
$4^{'}$& 0.093& 4.4&  0.90 &0.95 & 0.923 & 0.900    & $0.136 \pm 0.008 \pm 0.016$\\
$5^{'}$& 0.093& 4.2&  0.95 &1.05 & 0.982 & 0.934    & $0.090 \pm 0.008 \pm 0.010$\\
\end{tabular}
\label{tab:res0}
\end{center}
\end{table}

\begin{figure}
\centerline{\includegraphics[clip,width=1.0\textwidth]{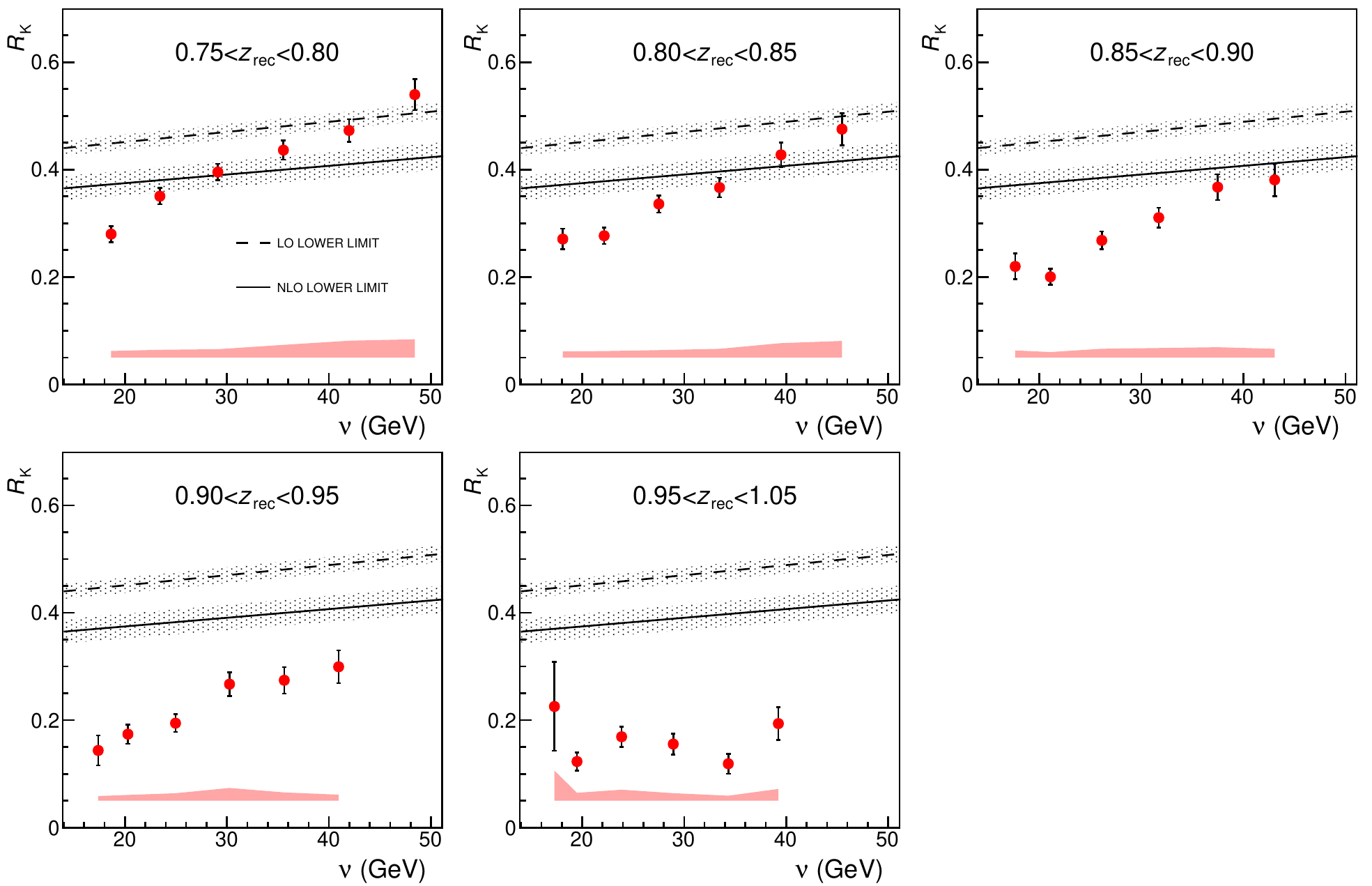}}
\caption{The K$^-$ over K$^+$ multiplicity ratio as a function of $\nu$ in bins of $z$, 
shown for the first bin in $x$.
The systematic uncertainties of the data points are
indicated by the shaded band at the bottom of each panel.
The shaded bands around the (N)LO lower limits
indicate their uncertainties. }

\label{fig:nu}
\end{figure}

In Fig.~\ref{fig:nu}, the dependence of $\rk$ on the virtual-photon energy 
$\nu$ in bins of the reconstructed $z$ variable is shown for the first $x$-bin. 
A clear $\nu$-dependence of $\rk$ is observed for all $z$-bins, except the last one.
Within experimental uncertainties, the observed dependence
on $\nu$ is linear and in the last bin a constant.
Note that at most 15\% of the observed variation of $\rk$ with $\nu$ can be explained by the fact 
that in a given $z$-bin events at different $\nu$ have somewhat different values of $x$ and $Q^2$.
The observed strong $\nu$ dependence suggests that for larger values of $\nu$ the ratio
$\rk$ is closer to the lower limit expected from pQCD than it is the case for smaller values of $\nu$. 
Numerical values for the $\nu$ dependence of $\rk$
in bins of $z_{\rm rec}$ are given for both $x$-bins in Ref. \cite{hepdata}.

In this analysis, the largest discrepancy between
pQCD expectations and experimental results is observed
in the region of large $z$ and small $y$, $i.e.$ small $\nu$. 
As exactly in this region the previously published COMPASS data 
\cite{comp_K} had shown the largest tension with the NLO pQCD fits of FFs, 
see Section \ref{sec:int}, 
the present results provide additional evidence that this tension 
is of physical origin.

\begin{figure}
\centerline{\includegraphics[clip,width=0.9\textwidth]{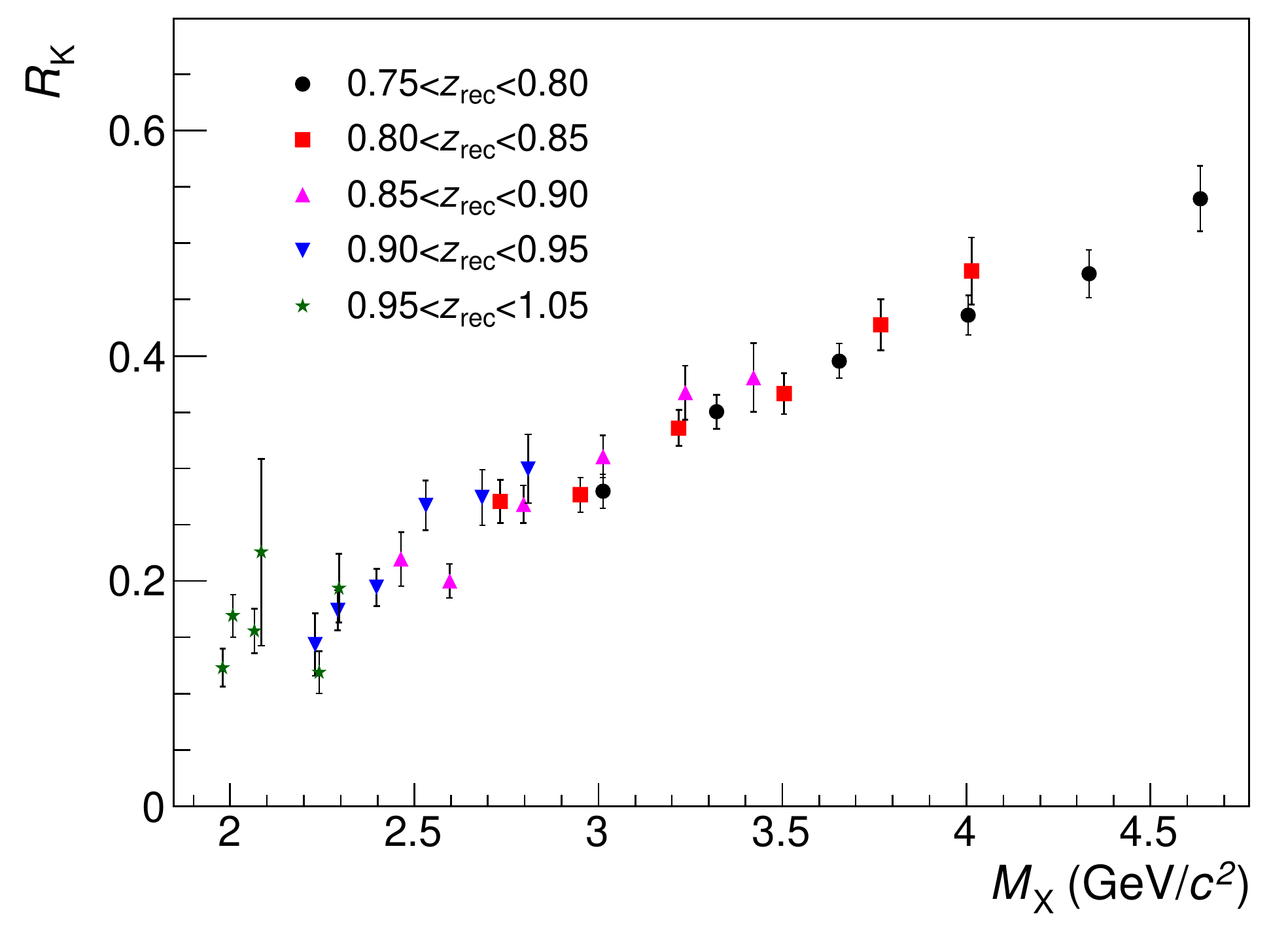}}
\caption{The K$^-$ over K$^+$ multiplicity ratio presented as a function of $M_X$. See text
for details.}
\label{fig:resmx}
\end{figure}

The observed violation of the pQCD expectations for the charged-kaon multiplicity 
ratio at large values of $z$ may be interpreted as follows. 
If the produced kaon carries a large fraction $z$ of the virtual-photon energy, 
there is only a small amount of energy left to fulfil conservation laws as 
$e.g.$ those for strangeness number and baryon number, 
which are not taken into account in the pQCD expressions for the SIDIS cross section. 
The larger the value of $z$, the smaller is the number of possible final states 
in the process under study. 
The natural variable to study the ``exclusivity'' of a process 
is the missing mass, which is approximately given by $M_{X}=\sqrt{M_{\rm p}^{2} + 2M_{\rm p} \nu(1-z) - Q^{2}(1-z)^2}$. 
As the factor $\nu(1-z)$ appears in the missing mass definition, both the $z$ and the $\nu$ dependence of $\rk$ may be described simultaneously by this variable.
Figure~\ref{fig:resmx} shows that 
$\rk$ as a function of $M_X$ follows a rather smooth behaviour. 
The disagreement between our data and the pQCD predictions suggests that a correction within the pQCD formalism is needed in order 
to take into account the phase space available for the hadronisation of the target remnant.
We observe that our data can be reconciled with the pQCD NLO prediction  
($\rk$ larger than about 0.4) only above the rather high $M_{X}$ value of about 4 GeV/$c^2$, 
which is rather surprising (see $e.g.$ Ref. \cite{dksw}).
Since the dominant term in $M_{X}$ is $\propto \sqrt{\nu(1-z)}$, this observation also suggests that for experiments with accessible values of $\nu$
smaller than those at COMPASS, the disagreement with pQCD calculations 
and possible deviations from these expectations may already be observed at smaller values of $z$.

\section{Summary}

In this Letter, the K$^-$ over K$^+$ multiplicity ratio $\rk$ measured 
in deep-inelastic kaon leptoproduction at large values of $z$ is presented for the first time. 
It is observed that the $\rk$ values fall below 
the lower limits calculated at LO and NLO accuracy in the pQCD formalism. 
In addition, we observe that the kaon multiplicity ratio $\rk$ strongly depends 
on the missing mass in the single-inclusive kaon production process.
Altogether, our observations suggest that more theory effort may 
be required in order to understand kaon production at high $z$. In particular, 
within the pQCD formalism an additional correction may be 
required that takes into account the phase space available for hadronisation.

\section*{Acknowledgements}
We would like to thank D. Stamenov for useful discussions.
We gratefully acknowledge the support of the CERN management and staff and the
skill and effort of the technicians of our collaborating institutes. 
This work was made possible by the financial support of our funding agencies.

\appendix

\section{Procedure for $z$-unfolding}

A typical unfolding procedure produces a covariance matrix
with non-negligible off-diagonal matrix elements.
These correlations are important and in many cases cannot be neglected,
as it is also emphasised in Ref. \cite{hermes}. In certain phenomenological 
analyses of published multiplicity data, however, these important pieces 
of information are erroneously neglected, which may lead to improper 
data treatment and thus to incorrect conclusions. In order to prevent 
such problems, we chose a simple unfolding method in our main analysis. 
We note that any correctly performed 
unfolding procedure can only decrease the value of $\rk$ 
measured at a given value of $z_{\rm rec}$, so that the choice of the unfolding 
procedure can not possibly explain the discrepancy observed between 
pQCD predictions and COMPASS results. 

As an example of a more sophisticated $z$-unfolding method, 
a procedure is presented that assures a smooth behaviour
of the resulting charged-kaon multiplicity ratio. Based
on MC data a smearing matrix is created, in which the 
probabilities are stored that the kaon with a generated value 
$z$ that belongs to a certain $z_{\rm gen}$-bin
is reconstructed in a certain $z_{\rm rec}$-bin.
The width of the $z$-bins is chosen to be 0.05 and
values of $z_{\rm rec}$ up to 1.10 are studied.
The obtained smearing matrix is given in Ref. \cite{hepdata} 
as supplemental material. In the next step, a functional
form for the K$^{\pm}$ multiplicities is assumed in the `true' 
phase space for data, which for MC data corresponds to the phase space 
of generated variables.
For the fit of the real data, the functional form  $\alpha \cdot \exp{(\beta z)}(1-z)^{\gamma}$ is used.
This function is integrated in bins of $z_{\rm gen}$, which are defined by the smearing matrix.
In this way, a vector of expectation values is obtained in the `true' 
phase space. This vector is multiplied by the smearing matrix, resulting
in expectation values for kaon yields in the reconstructed phase space.
The yield predictions obtained in this way are directly compared with the experimental values
by calculating a $\chi^2$ value. This value is minimised to find optimal parameters for the fitting function.
In order to obtain the uncertainty of the unfolded ratio,
the bootstrap method is used with 400 replicas of our data \cite{bootstrap}.
At a given value of $z$, the uncertainty of the ratio is taken
as Root Mean Square from the replicas distribution.
The effect of unfolding is rather small for all bins except the last one. The obtained results
are summarised in Table~\ref{tab:resunf} and the correlation matrix is given
in Table~\ref{tab:rescorr}.

\begin{table}[h!]
\begin{center}
\caption{The $z$-unfolded $\rk$ 
defined as $\int_{z_{\rm min}}^{z_{\rm max}} \frac{{\rm d}M^{{\rm K}^{-}}}{{\rm d}z} {\rm d}z / \int_{z_{\rm min}}^{z_{\rm max}}
\frac{{\rm d}M^{{\rm K}^{+}}}{{\rm d}z}{\rm d}z$, where $z_{\rm min(max)}$ denote bin limits in $z$. The data below 
(above) $x=0.05$ are presented in the top (bottom)
part of the table.} \label{tab:resunf}
\begin{tabular}{  c | c  c | c  }
bin  & $z_{\rm min}$   &  $z_{\rm max}$   &   $\rk \pm \delta R_{\rm K,\, stat.} \pm \delta R_{\rm K,\, syst.}$ \\ \hline
1 &  0.75 &0.80 &$0.416 \pm	 0.009 \pm	 0.018$       \\
2 &  0.80 &0.85 &$0.360 \pm	 0.010 \pm	 0.017$       \\
3 &  0.85 &0.90 &$0.289 \pm	 0.009 \pm	 0.014$       \\
4 &  0.90 &0.95 &$0.200 \pm	 0.014 \pm	 0.011$       \\
5 &  0.95 &1.00 &$0.085 \pm	 0.022 \pm	 0.007$       \\ \hline
$1^{'}$&  0.75 &0.80 &$0.237 \pm	 0.006 \pm	 0.011$       \\
$2^{'}$&  0.80 &0.85 &$0.202 \pm	 0.006 \pm	 0.010$       \\
$3^{'}$&  0.85 &0.90 &$0.165 \pm	 0.006 \pm	 0.009$       \\
$4^{'}$&  0.90 &0.95 &$0.123 \pm	 0.009 \pm	 0.007$       \\
$5^{'}$&  0.95 &1.00 &$0.068 \pm	 0.016 \pm	 0.005$       \\
\end{tabular}
\end{center}
\end{table}

\begin{table}[h!]
\begin{center}
\caption{The correlation matrix related to total uncertainties of the data presented in Table~\ref{tab:resunf}.}
\label{tab:rescorr}
\begin{tabular}{ c | c c c c c}

bin &  $1^{(')}$ & $2^{(')}$ & $3^{(')}$ & $4^{(')}$ & $5^{(')}$ \\ \hline
1 & 1.00 & 0.99 & 0.89 & 0.39 &-0.18 \\
2 & 0.99 & 1.00 & 0.94 & 0.47 &-0.12 \\
3 & 0.89 & 0.94 & 1.00 & 0.74 & 0.21 \\
4 & 0.39 & 0.47 & 0.74 & 1.00 & 0.81 \\
5 &-0.18 &-0.12 & 0.21 & 0.81 & 1.00 \\ \hline
$1^{'}$& 1.00 & 0.98 & 0.84 & 0.37 &-0.15 \\
$2^{'}$& 0.98 & 1.00 & 0.93 & 0.50 &-0.04 \\
$3^{'}$& 0.84 & 0.93 & 1.00 & 0.78 & 0.30 \\
$4^{'}$& 0.37 & 0.50 & 0.78 & 1.00 & 0.82 \\
$5^{'}$&-0.15 &-0.04 & 0.30 & 0.82 & 1.00 \\

\end{tabular}
\end{center}
\end{table}



\end{document}

%% file: Authors2018.tex
%
%
\section*{The COMPASS Collaboration}
\label{app:collab}
\renewcommand\labelenumi{\textsuperscript{\theenumi}~}
\renewcommand\theenumi{\arabic{enumi}}
\begin{flushleft}
R.~Akhunzyanov\Irefn{dubna}, 
M.G.~Alexeev\Irefn{turin_u},
G.D.~Alexeev\Irefn{dubna}, 
A.~Amoroso\Irefnn{turin_u}{turin_i},
V.~Andrieux\Irefnn{illinois}{saclay},
N.V.~Anfimov\Irefn{dubna}, 
V.~Anosov\Irefn{dubna}, 
A.~Antoshkin\Irefn{dubna}, 
K.~Augsten\Irefnn{dubna}{praguectu}, 
W.~Augustyniak\Irefn{warsaw}
% ...
\end{flushleft}
%
%
\begin{Authlist}
\item \Idef{aveiro}{University of Aveiro, Dept.\ of Physics, 3810-193 Aveiro, Portugal}
\item \Idef{bochum}{Universit\"at Bochum, Institut f\"ur Experimentalphysik, 44780 Bochum, Germany\Arefs{l}\Aref{s}}
\item \Idef{bonniskp}{Universit\"at Bonn, Helmholtz-Institut f\"ur  Strahlen- und Kernphysik, 53115 Bonn, Germany\Arefs{l}}
\item \Idef{bonnpi}{Universit\"at Bonn, Physikalisches Institut, 53115 Bonn, Germany\Arefs{l}}
\item \Idef{brno}{Institute of Scientific Instruments, AS CR, 61264 Brno, Czech Republic\Arefs{m}}
\item \Idef{calcutta}{Matrivani Institute of Experimental Research \& Education, Calcutta-700 030, India\Arefs{n}}
\item \Idef{dubna}{Joint Institute for Nuclear Research, 141980 Dubna, Moscow region, Russia\Arefs{o}}
\item \Idef{freiburg}{Universit\"at Freiburg, Physikalisches Institut, 79104 Freiburg, Germany\Arefs{l}\Aref{s}}
\item \Idef{cern}{CERN, 1211 Geneva 23, Switzerland}
\item \Idef{liberec}{Technical University in Liberec, 46117 Liberec, Czech Republic\Arefs{m}}
\item \Idef{lisbon}{LIP, 1000-149 Lisbon, Portugal\Arefs{p}}
\item \Idef{mainz}{Universit\"at Mainz, Institut f\"ur Kernphysik, 55099 Mainz, Germany\Arefs{l}}
\item \Idef{miyazaki}{University of Miyazaki, Miyazaki 889-2192, Japan\Arefs{q}}
\item \Idef{moscowlpi}{Lebedev Physical Institute, 119991 Moscow, Russia}
\item \Idef{munichtu}{Technische Universit\"at M\"unchen, Physik Dept., 85748 Garching, Germany\Arefs{l}\Aref{r}}
\item \Idef{nagoya}{Nagoya University, 464 Nagoya, Japan\Arefs{q}}
\item \Idef{praguecu}{Charles University in Prague, Faculty of Mathematics and Physics, 18000 Prague, Czech Republic\Arefs{m}}
\item \Idef{praguectu}{Czech Technical University in Prague, 16636 Prague, Czech Republic\Arefs{m}}
\item \Idef{protvino}{State Scientific Center Institute for High Energy Physics of National Research Center `Kurchatov Institute', 142281 Protvino, Russia}
\item \Idef{saclay}{IRFU, CEA, Universit\'e Paris-Saclay, 91191 Gif-sur-Yvette, France\Arefs{s}}
\item \Idef{taipei}{Academia Sinica, Institute of Physics, Taipei 11529, Taiwan\Arefs{tw}}
\item \Idef{telaviv}{Tel Aviv University, School of Physics and Astronomy, 69978 Tel Aviv, Israel\Arefs{t}}
\item \Idef{triest_u}{University of Trieste, Dept.\ of Physics, 34127 Trieste, Italy}
\item \Idef{triest_i}{Trieste Section of INFN, 34127 Trieste, Italy}
\item \Idef{turin_u}{University of Turin, Dept.\ of Physics, 10125 Turin, Italy}
\item \Idef{turin_i}{Torino Section of INFN, 10125 Turin, Italy}
\item \Idef{tomsk}{Tomsk Polytechnic University,634050 Tomsk, Russia\Arefs{nauka}}
\item \Idef{illinois}{University of Illinois at Urbana-Champaign, Dept.\ of Physics, Urbana, IL 61801-3080, USA\Arefs{nsf}}
\item \Idef{warsaw}{National Centre for Nuclear Research, 00-681 Warsaw, Poland\Arefs{u}}
\item \Idef{warsawu}{University of Warsaw, Faculty of Physics, 02-093 Warsaw, Poland\Arefs{u}}
\item \Idef{warsawtu}{Warsaw University of Technology, Institute of Radioelectronics, 00-665 Warsaw, Poland\Arefs{u} }
\item \Idef{yamagata}{Yamagata University, Yamagata 992-8510, Japan\Arefs{q} }
\end{Authlist}
%
%
\renewcommand\theenumi{\alph{enumi}}
\begin{Authlist}
\item [{\makebox[2mm][l]{\textsuperscript{\#}}}] Corresponding authors
\item [{\makebox[2mm][l]{\textsuperscript{*}}}] Deceased
\item \Adef{a}{Also at Instituto Superior T\'ecnico, Universidade de Lisboa, Lisbon, Portugal}
\item \Adef{b}{Also at Dept.\ of Physics, Pusan National University, Busan 609-735, Republic of Korea and at Physics Dept., Brookhaven National Laboratory, Upton, NY 11973, USA}
\item \Adef{ictp}{Also at Abdus Salam ICTP, 34151 Trieste, Italy}
\item \Adef{r}{Supported by the DFG cluster of excellence `Origin and Structure of the Universe' (www.universe-cluster.de) (Germany)}
\item \Adef{d}{Also at Chubu University, Kasugai, Aichi 487-8501, Japan\Arefs{q}}
\item \Adef{x}{Also at Dept.\ of Physics, National Central University, 300 Jhongda Road, Jhongli 32001, Taiwan}
\item \Adef{e}{Also at KEK, 1-1 Oho, Tsukuba, Ibaraki 305-0801, Japan}
\item \Adef{g}{Also at Moscow Institute of Physics and Technology, Moscow Region, 141700, Russia}
\item \Adef{h}{Present address: RWTH Aachen University, III.\ Physikalisches Institut, 52056 Aachen, Germany}
\item \Adef{yerevan}{Also at Yerevan Physics Institute, Alikhanian Br. Street, Yerevan, Armenia, 0036}
\item \Adef{y}{Also at Dept.\ of Physics, National Kaohsiung Normal University, Kaohsiung County 824, Taiwan}
\item \Adef{turin_p}{Also at University of Eastern Piedmont, 15100 Alessandria, Italy}
\item \Adef{c}{    Supported by the DFG Research Training Group Programmes 1102 and 2044 (Germany)} 
%
%
\item \Adef{l}{    Supported by BMBF - Bundesministerium f\"ur Bildung und Forschung (Germany)}
\item \Adef{s}{    Supported by FP7, HadronPhysics3, Grant 283286 (European Union)}
\item \Adef{m}{    Supported by MEYS, Grant LG13031 (Czech Republic)}
\item \Adef{n}{    Supported by B.Sen fund (India)}
\item \Adef{o}{    Supported by CERN-RFBR Grant 12-02-91500}
\item \Adef{p}{\raggedright 
                   Supported by FCT - Funda\c{c}\~{a}o para a Ci\^{e}ncia e Tecnologia, COMPETE and QREN, Grants CERN/FP 116376/2010, 123600/2011 
                   and CERN/FIS-NUC/0017/2015 (Portugal)}
\item \Adef{q}{    Supported by MEXT and JSPS, Grants 18002006, 20540299, 18540281 and 26247032, the Daiko and Yamada Foundations (Japan)}
\item \Adef{tw}{   Supported by the Ministry of Science and Technology (Taiwan)}
\item \Adef{t}{    Supported by the Israel Academy of Sciences and Humanities (Israel)}
\item \Adef{nauka}{Supported by the Russian Federation  program ``Nauka'' (Contract No. 0.1764.GZB.2017) (Russia)}
\item \Adef{nsf}{  Supported by the National Science Foundation, Grant no. PHY-1506416 (USA)}
\item \Adef{u}{    Supported by NCN, Grant 2015/18/M/ST2/00550 (Poland)}
\end{Authlist}